# Harry Potter's Cloak

X. F. Zhu[†], B. Liang[†], J. Tu[†], D. Zhang[·,*], J. C. Cheng[*]

Key Laboratory of Modern Acoustics, MOE, Institute of Acoustics,

Department of Physics, Nanjing University

Nanjing 210093, P. R. China

[†]These authors contributed equally to this work.

[*] To whom correspondence should be addressed. Emails:

jccheng@nju.edu.cn (J. C. C.)

dzhang@nju.edu.cn (D.Z.)

**Abstract:**

The magic "Harry Potter's cloak" has been the dream of human beings for really long time. Recently, transformation optics inspired from the advent of metamaterials offers great versatility for manipulating wave propagation at will to create amazing illusion effects. In the present work, we proposed a novel transformation recipe, in which the cloaking shell somehow behaves like a "cloaking lens", to provide almost all desired features one can expect for a real magic cloak. The most exciting feature of the current recipe is that an object with arbitrary characteristics (e.g., size, shape or material properties) can be invisibilized perfectly with positive-index materials, which significantly benefits the practical realization of a broad-band cloaking device fabricated with existing materials. Moreover, the one concealed in the hidden region is able to undistortedly communicate with the surrounding world, while the lens-like cloaking shell will protect the cloaked source/sensor from being traced back by outside detectors by creating a virtual image.

A continuously changing refractive index results in changes of light propagation to create illusion effect. The most common example is the curved glass lens that can redirect the electromagnetic field by either bending or focusing the light beam. With conventional homogeneous materials, optical design is mainly a matter of determining the interface between two materials. The advent of artificial complex metamaterials enables novel wave manipulations that are impossible to achieve with the nature materials (1, 2, 3). Transformation optics, introduced with the great progress of metamaterial implementation, establishes a correspondence between coordinate transformations and material constitutive properties and inspires researchers to take a fresh look at the conventional foundations of optics (1). Analogous to general relativity, the principle of transformation optics shows that the field of wave (light, acoustic waves, and matter wave, etc.) can also be controlled in an arbitrary manner, with specially designed structures (4). The ability of designing and engineering the wave field offers researchers great versatility to manipulate wave propagation at will, and then create amazing illusion effects in which the invisibility-cloaking is perhaps the most popular and intriguing topic (5, 6, 7).

So far, a wide variety of approaches have been proposed for cloaking purpose.

"Push-forward mapping" (5, 8, 9), one of the most common strategies, was developed to exclude the wave from the inner domain without perturbing the exterior fields by expanding a point into a hole where the cloaked object would be hidden. However, owing to the existence of singularity in the coordinate transformation, infinite parameters would be required for the cloaking materials, which limited the application of this approach in a narrow bandwidth. Moreover, the object concealed with this method can not "see" the outside world, which leads to the problem of "double-blind". By sacrificing some degree of cloaking effects, other approaches (6, 10, 11) were then developed to achieve wider operating bandwidth using the materials with less extreme parameters, whereas the modified recipes offered little help to the double-blind problem. Therefore, another approach, so-called "external cloaking" (12, 13, 14), was proposed to overcome this problem, in which the cloaked object could be "canceled out" by its "anti-object" and be able to share information with the surrounding. However, the external cloak can only work in a steady state, since the multiple-scattering between the cloaked object and anti-object has to reach equilibrium to achieve the destructive interference of scattered waves. Particularly, in order to serve as an anti-object to the cloaked one, the cloaking material must be "custom-made" with double-negative parameters, which creates an obstacle for the practical realization of the external cloaking. Furthermore, although the cloaked object is not "blinded", one can only catch anamorphic images for the outside world, since the penetrating waveform could be distorted inside the cloaked region. Recently, Zhu et al. (15) proposed another transformation model, named as "magnifying superlens", to fulfill the task of "non-double-blinded" cloaking using complementary materials designed with single-negative parameters, which are not dependent on the properties of the host material and cloaked object. Although the application of single-negative complementary materials will reduce the fabrication difficulties, the broad-band cloaking still can hardly be achieved because the operating frequencies of complementary media must overlap each other. In addition, this method can only be applied to the cloaked object that is more compressible than the host medium, since the cloaked object should be mapped to a bulk of host medium with relatively larger size.

Despite of the fact that various transformation models have been developed, the real "Harry Potter's cloak" is still out of reach. In the present work, a novel transformation model is proposed to unprecedentedly approach the dream cloak. This method moves from a "superlens" to a "cloaking lens", without referring to the complementary concept. The invisibility cloak designed with current transformation model would allow one to see the outside world un-anamorphically, while perfectly concealing his/her presence. This model should be fit for cloaking objects with arbitrary characteristics (e.g., size, shape or material properties) under all circumstances. Most important of all, the current scheme allows designing the cloaking shell with only positive-index materials, which dramatically relaxes the extreme property requirements for the cloak material and offers the designer great possibility to experimentally realize the cloaking structure with existing materials within a broad frequency band.

The principle of current transformation model is general and equally applicable to various wave forms (e.g., acoustic waves, electromagnetic waves, matter waves), as long as the wave equations can keep invariant under coordinate transformations. For an arbitrary wave form, the scalar Helmholtz equation with a source term can be written as (16)

$$\frac{1}{\sqrt{|\boldsymbol{g}|}} \sum_{i,j=1}^{n} \frac{\partial}{\partial x_i}\left(\sqrt{|\boldsymbol{g}|} g^{ij} \frac{\partial p}{\partial x_j}\right) + k^2 p = f, \tag{1}$$

for a Riemannian metric $\boldsymbol{g} = (g_{ij})$ in $n$-dimensional space, where $|\boldsymbol{g}| = \det(g_{ij})$, $(g^{ij}) = \boldsymbol{g}^{-1} = (g_{ij})^{-1}$. Here, we only consider the simplest nontrivial 2D situation in acoustics, where $\omega = k$, $\boldsymbol{\rho}^{-1} = (\rho_0^{-1}\delta_{ij})(\sqrt{|\boldsymbol{g}|} g^{ij})$, $\boldsymbol{\kappa} = (\kappa_0 \delta_{ij})(\sqrt{|\boldsymbol{g}|})^{-1}$ with $\rho_0^{-1}\delta_{ij}$ and $\kappa_0 \delta_{ij}$ being the inverse mass density tensor and bulk modulus tensor in the virtual space. For the mapping between orthogonal coordinates in virtual space $\boldsymbol{x}$ and real space $\boldsymbol{x}'$, the corresponding metric tensor is

$$\boldsymbol{g} = \mathrm{diag}\left\{\left[\frac{h_{x_1}}{h_{x_1'}} \frac{\partial x_1}{\partial x_1'}\right]^2, \left[\frac{h_{x_2}}{h_{x_2'}} \frac{\partial x_2}{\partial x_2'}\right]^2\right\}, \tag{2}$$

where $h_{x_1}$, $h_{x_2}$ and $h_{x_1'}$, $h_{x_2'}$ are the scale factors in the orthogonal coordinates. As illustrated in Fig. 1, an equivalence is established in the 2D cylindrical coordinates between a bulk of host medium A $(0 < r < a_1)$ and the cloaked object A' $(0 < r' < a_2)$, so does an annulus of host medium B $(a_1 < r < b)$ and the cloak shell B' $(a_2 < r' < b)$. Here $r$ (or $r'$) refers to the radius of studied region. The effective mass density $\rho_1$ of the cloaked object should be assumed to be the same as the mass density $\rho_0$ of the host medium, which then has practical significance for cloaking the object buoyant in fluid. For simplicity we choose the circle configuration for the cloaked object. The mapping relationship between virtual space $x$ and real space $x'$ can be written as

$$r' = f(r) = \begin{cases} r(a_2/a_1), & 0 < r < a_1 \\ (rb^{n-1})^{1/n}, & a_1 < r < b \\ r, & r > b \end{cases}, \quad \vartheta' = \vartheta, \tag{3}$$

where the mapping order $n \in (0,1) \cup (1,\infty)$, and $b = (a_2^n/a_1)^{1/(n-1)}$. Suppose the mass density tensor and the bulk modulus tensor of the cloaked object are $\boldsymbol{\rho}_{\text{ob}} = \rho_0 \delta_{ij}$ and $\boldsymbol{\kappa}_{\text{ob}} = \kappa_1 \delta_{ij}$, respectively. According to Eqs. (1)-(3), for particular $a_2$ (the radius of the cloaked object) and $n$ (the mapping order), the mass density tensor $\boldsymbol{\rho}_{\text{S}}$, the bulk modulus tensor $\boldsymbol{\kappa}_{\text{S}}$, and the external radius of the cloaking shell $b$ can be readily determined as:

$$\boldsymbol{\rho}_{\text{S}} = \rho_0 (n\delta_{i1} + n^{-1}\delta_{i2})\delta_{ij}, \quad \boldsymbol{\kappa}_{\text{S}} = (a_2/r')^{2n-2} \kappa_1/(n\kappa_0)\delta_{ij}, \quad b = a_2(\kappa_1/\kappa_0)^{1/(2n-2)}, \tag{4}$$

where $\delta_{ij}$ is Kronecker delta and $r' \in (a_2, b)$. It can be apparently observed that only positive parameters are required for the cloaking materials, which is quite desirable for overcoming the hurdle in the practical implementation of cloaking device in a broad bandwidth. Normally, the cloaking material parameters designed using previously proposed transformation strategies are unique to the properties of cloaked objects and the host medium (12, 13, 14), which will significantly limit the experimental implementation of the invisibility-cloaking. The present transformation structure offers high flexibility for engineering the cloaking materials by adjusting the structural parameters of the shell. The strict demand on the cloaking shell properties could be reduced by increasing the shell thickness. According to Eq. (4), when the mapping order $n$ goes lower (*viz.* thicker cloaking shell), the material of the shell tends to be less anisotropic and inhomogeneous, which provides higher degrees of freedom for designing the cloaking shell.

As shown in Fig. 1, it is apparent that $b > a_2$. Since $b = (a_2^n/a_1)^{1/(n-1)} = a_2(\kappa_1/\kappa_0)^{1/(2n-2)}$, two cases should be considered individually for $a_1 > a_2$, $\kappa_1 < \kappa_0$ as $n \in (0,1)$ and $a_1 < a_2$, $\kappa_1 > \kappa_0$ as $n \in (1,\infty)$. Our transformation scheme is demonstrated in the Fig. 2. We first consider the case that the cloaked object is equivalent to a larger bulk of host medium, i.e. $a_1 > a_2$ for $n = (0,1)$. As mentioned above, it is obvious that the bulk modulus of the cloaked object is smaller than that of the host medium, which means the cloaked object is more compressible than the host medium. The coordinates of physical space (Fig. 2B) are curved transformations of Cartesian coordinates in a virtual space (Fig. 2A), in which the waves propagate along straight lines. The corresponding full-wave simulations for the cloaking effects of the designed devices are illustrated in the Supporting Online Materials (17). As shown in Fig. 2B, after applying current transformation, the cloaking shell serves as a special lens to bend the incident waves, then guides them to flow around the cloaked object inside the cloaking shell (e.g., yellow lines) or straightly penetrate the hidden region (e.g., blue lines), finally returns them to their original paths on the far side, which makes the outside observers be unaware of the contents in the hidden region. In Fig. 2B, it is notable that the penetrated waves inside the cloaked region are also following straight paths. This feature not only allows the transformation devise to solve the double-blinded problem as the external cloaking does, but also ensures the cloaked one to undistortedly share the information with outside world, which might offer a better strategy to design a genuine "Harry Potter's cloak". There is a special case in this category: the thickness of the cloaking shell has a minimum (*viz.* $b = a_1$). Under this condition, the shell is mapped to a circle line after the transformation, which leads to the concept of superlens cloaking given by Zhu et al. (15)

Then we consider the case that the bulk modulus of the cloaked object is larger than that of the host medium, i.e., the cloaked object is less compressible than the host medium. With the current transformation scheme, the cloaked object is equivalent to a smaller bulk of host media, i.e. $a_1 < a_2$ for $n = (1, \infty)$. The one-to-one correspondences between the virtual and physical spaces for this case are illustrated in Figs. 2C and 2D, respectively. It is obvious that, similar to the case demonstrated in Figs. 2A and 2B, the cloaking structure designed with current transformation scheme can conceal the object perfectly and enable undistorted information exchange between the cloaked objects and the surrounding environment, although the wave rays propagating inside the cloaking shell would be bent apart from the hidden region due to the material compressibility difference. There also exists a special case in this category: when the cloaked object is extremely incompressible (nearly a rigid body), it can be mapped to a point-like host medium after the transformation $(a_1 \to 0)$. This kind of rigid body can be well hidden using the proposed transformation strategy and the cloaking material parameters designed are consistent with those given by Pendry et al. (5)

Going far beyond "non-double-blinded" cloaking, current transformation structure also owes another promising feature, which allows the one inside the cloaked region to transmit useful information out without being traced back. Imagine we put a source (*O*) in the cloaked region (see Fig. 3). Now the cloaking shell also behaves like a lens. Before entering the outside space, the wave rays sent from the source will be refracted twice on the internal (red circle) and external (dark cyan circle) interfaces of the shell. Due to the "lens refraction", a virtual image ($O^*$), whose size and position are different from the real image (*O*), is created in the hidden region, so that the real source can be protected from being detected by outside observers.

The current transformation structure, which could be regarded as building a bridge between the conventional lens and the transformation optics, realizes the purpose of "seeing without being seen" and unprecedentedly gets closer to the magic Harry Potter's cloak. The designed cloaking shell parameters are not strictly subjected to the characteristics of the concealed object, neither are suffering from the demand of extreme properties (e.g., singularity or negative-index). The ability of designing the cloaking shell with positive parameters will offer great freedom for the practical fabrication and implementation of invisibility-cloaking within a broadband spectrum. Furthermore, this recipe enables the source inside the hidden region to freely transmit signals out without exposing its actual information (e.g., size and position).

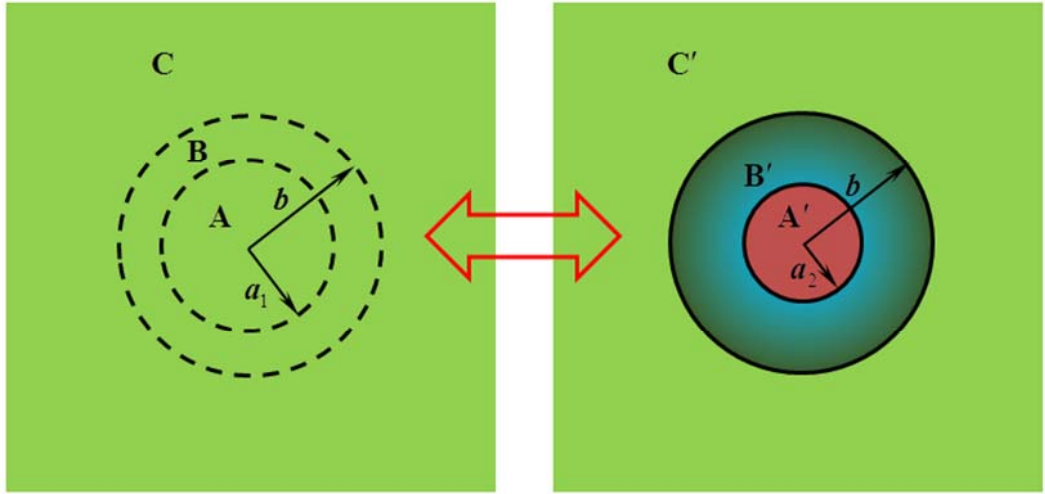

Fig. 1. The schematic illustration of current transformation strategy. The device performs a coordinate transformation from the virtual space (left) to physical space (right). The virtual space is empty and flat. In the physical space, an object is concealed in the cloaked region without being "blinded". The mapping approach is straightforward, where $A \to A'$, $B \to B'$, and $C \to C'$. $A'$ is filled with the cloaked object ($0 < r' < a_2$) and $B'$ is filled with the cloaking material ($a_2 < r' < b$).

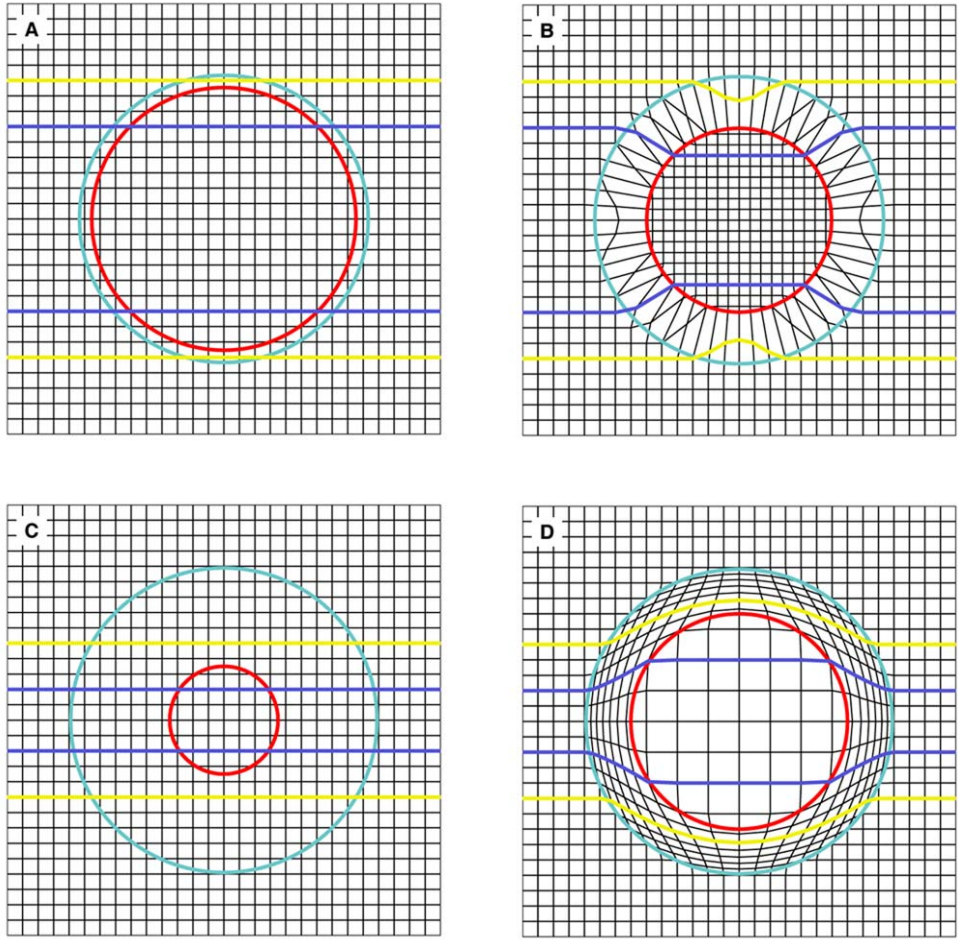

Fig. 2. Coordinate transformations of the "non-blinded" cloaking device in two cases, in which the cloaked object is either more compressible (**A** and **B**) or less compressible (**C** and **D**) than the host medium. In the virtual space (**A** and **C**), the device creates the illusion that waves (yellow and blue lines) propagate through a space that is empty and flat. The virtual plane carries the coordinate grid that is mapped onto the physical space, where the incident wave beams are bent convergently (**B**) or divergently (**D**) by the cloaking shell. The waves seem to pass through an empty area if observed in the far field. The yellow lines are steered around the cloaked object within the cloaking shell, while the blue lines penetrate through the cloaked object without phase distortion.

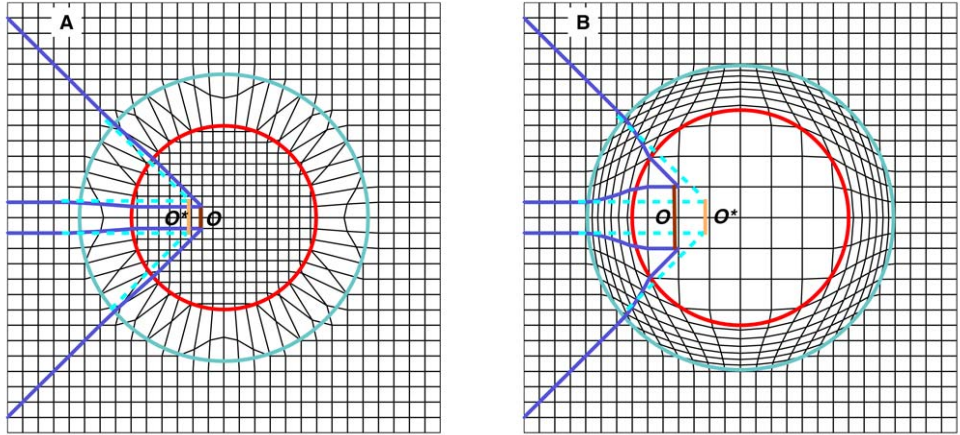

Fig. 3. The invisible transmitter. A transmitter inside the hidden region can freely interrogate the outside world. The cloaking shell, acting like a lens, can protect the signal source (*O*) against outside detection by creating a virtual image (*O\**). **A**: The virtual image looks larger than the original one, when the cloaked object is more compressible than the host medium. **B**: The virtual image looks smaller than the original source, when the cloaked object is less compressible than the host medium.

## Supporting Online Materials

### 1. Full-wave simulations for acoustic wave

Full-wave simulations were performed using finite element method (FEM) to demonstrate the cloaking properties of the designed structure. The following figures illustrate the simulation results for the cases of $a_1 > a_2$, $\kappa_1 < \kappa_0$ as $n \in (0,1)$ and $a_1 < a_2$, $\kappa_1 > \kappa_0$ as $n \in (1,\infty)$, respectively.

We first investigate the case of a scatterer illuminated by plane wave (SFig. 1). Here, the wavelength of the incident wave is assumed to be $\lambda = 0.5$ m. The host medium is set to be water with the bulk modulus and the mass density to be $\kappa_0 = 2.19$ GPa and $\rho_0 = 998$ kg/m$^3$, respectively. The pressure field distributions for the scatterer more (less) compressible than the host medium are illustrated in SFigs.1 A and B (SFigs.1 C and D). As shown in SFigs. A and C, the plane waves are strongly disturbed by the bare scatterers, which results in the backward reflection and sharp-edged shadow. SFigures B and D show the cloaking effect, where the low-reflection and shadow-reducing properties are clearly demonstrated. The plane wave field is undisturbed outside the cloaking shell. More important, the plane wave can penetrate through the cloaked object without changing the shape of wavefront, which makes it capable to undistortedly receive information from the outside. Another notable point is that the resonance phenomenon, which is claimed to be the particular characteristic effect occurring on the interface between the complementary media with negative parameters, is successfully suppressed by the use of the positive-index materials.

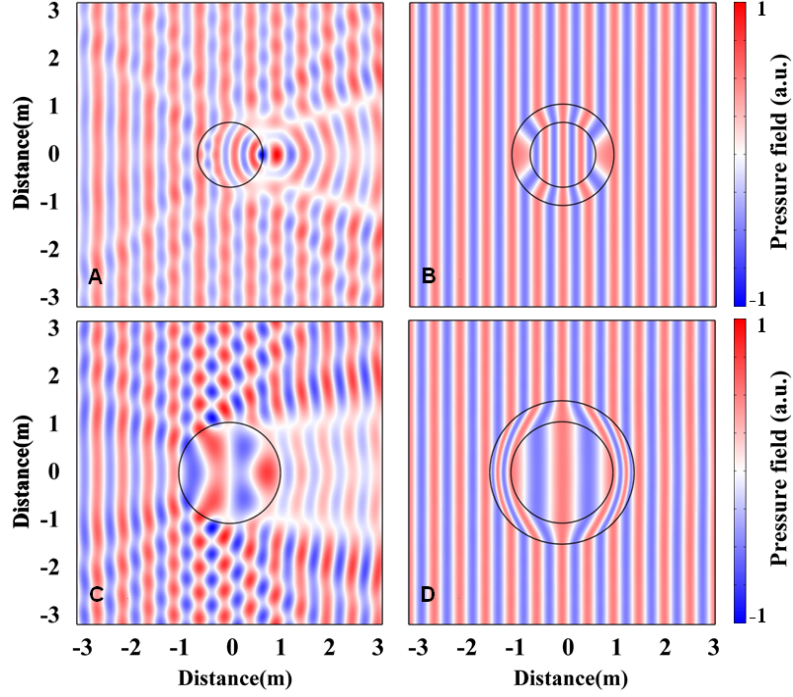

SFig. 1. The pressure field distribution for acoustic wave incident from a plane source. In **A** and **B**, the scattering object which is more compressible than the host medium is illuminated without/with a cloak. The bulk modulus, the mass density, and the radius of the cloaked object are $\kappa_{ob} = 0.49\kappa_0$, $\rho_{ob} = \rho_0$, and $a_2 = 0.64$ m, respectively. The outer radius of the cloak is readily determined as $b = 1$m for the mapping order $n = 0.2$. In **C** and **D**, an object which is less compressible than the host medium is illuminated without/with a cloak. The bulk modulus, the mass density and the radius of the cloaked object are $\kappa_{ob} = 4\kappa_0$, $\rho_{ob} = \rho_0$, and $a_2 = 1$m, respectively. The outer radius of the cloak is readily determined as $b = \sqrt{2}$ m for $n = 3$. In all cases, the acoustic wavelength in the host medium is $\lambda = 0.5$ m.

The full-wave simulations were also performed for the case of a point source (S) inside the cloaked region. SFigures 2 A and B (C and D) illustrated the pressure field distributions for the cloaked medium is more (less) compressible than the host one.

According to the comparison between SFigs. 2 A(C) and B(D), the utilization of specially designed cloak shell can eliminate the wave field disturbance caused by the presence of the scatterer. This feature enables the highest flexibility and stability for the source to communicate with outside world. On the other hand, it is impossible for the outside observer to trace back the original position of the source, since the wavefront inside the cloaked region is bent with the "cloaking lens" shell.

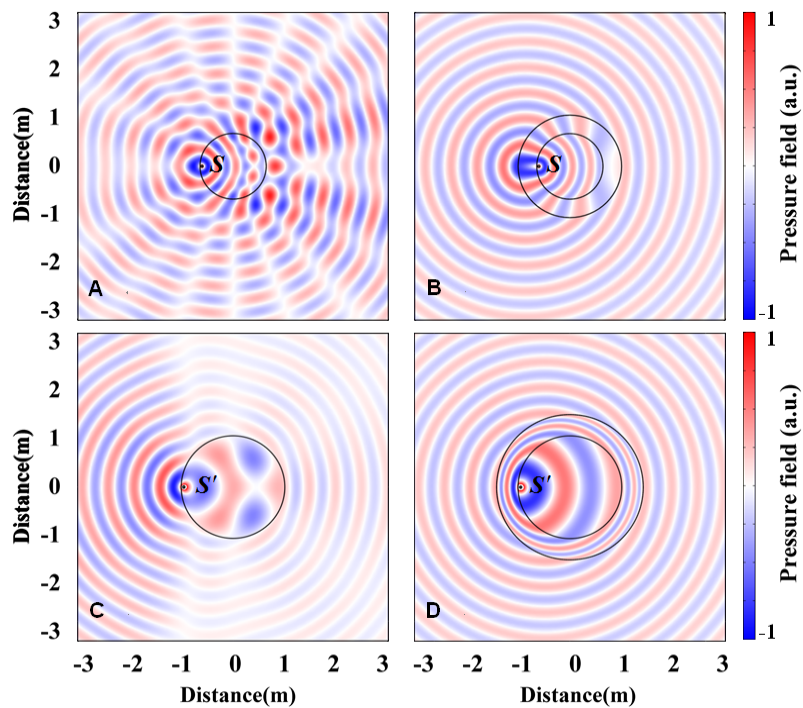

SFig. 2. The pressure field distribution for acoustic wave sending from a point source inside the cloaked region, without (**A** and **C**) or with (**B** and **D**) the cloaking shell. In **A** and **B**, the cloaked object is more compressible than the host medium and the point source $S$ located at $(-0.64\,\text{m}, 0)$. In **C** and **D**, the cloaked object is less compressible than the host medium, and $S'$ located at $(-1\,\text{m}, 0)$. In all cases, the acoustic wavelength in the host medium is $\lambda = 0.5\,\text{m}$.

## 2. Full-wave simulations for electromagnetic wave

It has been mentioned in the main body of the manuscript that the current transformation scheme can be equally applied to various wave forms (e.g., acoustic waves, electromagnetic waves, matter waves). Here, the full-wave simulations were also carried out for the scatterers illuminated with electromagnetic plane waves. The comparison between SFigs. 3 A (C) and B (D) shows the cloaking effect for the object whose permittivity is larger (smaller) than that of the host medium. It is obvious that the pressure field will be disturbed due to the presence of a bare scatterer (SFigs. 3 A and C). Bent with the "cloaking lens" shell (see SFigs. 3 B and D), the distorted waves can be guided back to their original paths. Therefore, in the eyes of observers, the incident waves seem to pass through an empty area and the scatterer is concealed perfectly. It notable that the incident waves can straightly penetrate the cloaked region, which enables the one inside the hidden region to communicate with surrounds undistortedly.

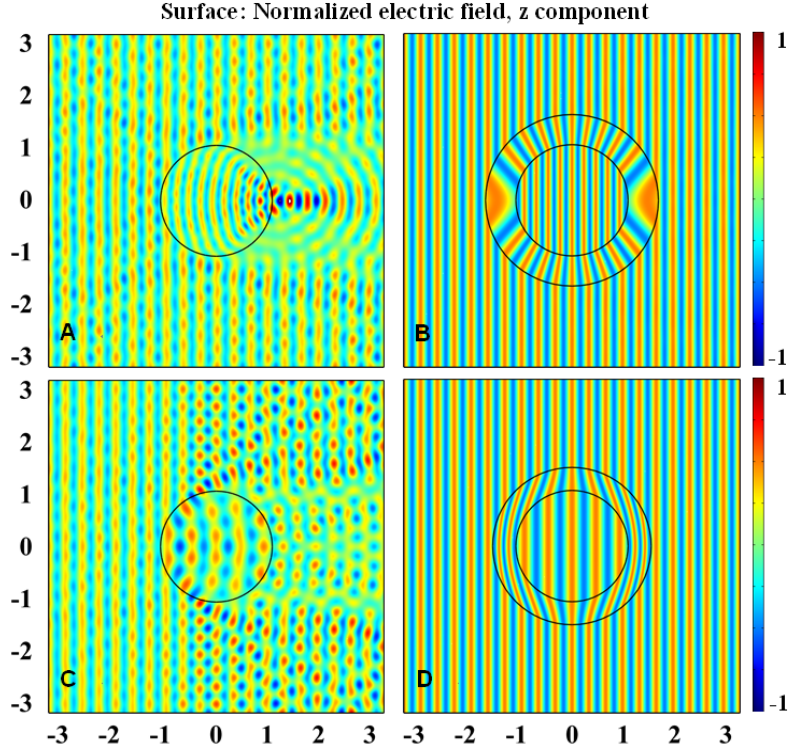

SFig. 3. Normalized electric field distribution for TE wave incident from a plane source. A bare scatterer is shielded by nothing [**A**, **C**] or a cloak [**B**, **D**]. For **A** and **B**, the material parameters of host medium and cloaked object are $\varepsilon_h = 1$, $\mu_h = 1$ and $\varepsilon_o = 2$, $\mu_o = 1$, respectively. For **C** and **D**, the material parameters of host matrix and cloaked object are $\varepsilon_h = 2$, $\mu_h = 1$ and $\varepsilon_o = 1$, $\mu_o = 1$, respectively. In **B**, the structural parameters are $a_2 = 1$, $b = 1.5422$, and the mapping order is $n = 0.2$; in **D**, the corresponding parameters are: $a_2 = 1$, $b = 1.1225$, and $n = 2$. In all cases, the wavelength in the host medium is $\lambda = 0.3$ unit.